\documentclass[aps,twocolumn,superscriptaddress]{revtex4}

\usepackage{amsmath,amssymb,bm}
\usepackage{graphicx}
\usepackage{amsfonts}
\usepackage{color}
\usepackage{color}

\graphicspath{{./figs/}}

\newcommand{\nn}{\nonumber}                                           
\newcommand{\va}[1]{\langle{#1}\rangle}                               

\newcommand{\ga}{\gamma}

\newcommand{\ro}{\rho}

\newcommand{\Tr}{\mathop{\rm Tr}\nolimits}

\begin{document}

\thispagestyle{empty}
\date{\today}
\preprint{\hbox{...}}

\title{Exotic glueball $0^{\pm -}$ states in QCD Sum Rules}

 \author{Alexandr Pimikov}
\email{pimikov@mail.ru}
\affiliation{Institute of Modern
    Physics, Chinese Academy of Science, Lanzhou 730000, China}

\author{Hee-Jung Lee}
\affiliation{Department of Physics Education, Chungbuk National
    University, Cheongju, Chungbuk 28644, Korea}

\author{Nikolai Kochelev}
\email{kochelev@theor.jinr.ru}
\affiliation{Institute of Modern Physics,
    Chinese Academy of Science, Lanzhou 730000, China}

\affiliation{Bogoliubov Laboratory of Theoretical Physics, Joint
    Institute for Nuclear Research,\\ Dubna, Moscow Region, 141980
    Russia}

\author{Pengming Zhang}
\affiliation{Institute of Modern Physics,
    Chinese Academy of Science, Lanzhou 730000, China}

\author{Viachaslau Khandramai}
\affiliation{International Center for Advanced Studies, Technical University, Gomel, 246746, Belarus}

\begin{abstract}
The lowest dimension three-gluon currents
that couple to the exotic $0^{\pm-}$ glueballs
have been constructed using the helicity formalism.
Based on the constructed currents,
we obtain new QCD SRs that have been used to extract the masses and the decay constants of
the scalar exotic $0^{\pm-}$ glueballs.
We estimate the masses for the scalar state and for the pseudoscalar state to be
${m_+=9.8^{+1.3}_{-1.4}}$~GeV and 
$m_-=6.8^{+1.1}_{-1.2}$~GeV.


\end{abstract}

\maketitle

\section{Introduction}

Glueballs are composite particles that contain gluons and no valence quarks.
Theoretically, glueball should exist because of the non-Abelian and confinement properties of 
Quantum Chromodynamics (QCD),
 due to the gluon self-interaction and strong ``dressing'' through vacuum fluctuations.
However, there is no clear experimental evidence and glueballs remain undiscovered~\cite{Jia:2016cgl}.
Their mixing with ordinary meson states makes it difficult to discover glueballs in an experimental search.
Glueball studies are important for phenomenology both at the running
and projected large-scale experiments in many research centers:
Belle (Japan),  BESIII (Beijing, China),
 LHC (CERN), GlueX (JLAB,USA),
NICA (Dubna, Russia), HIAF (China) and FAIR (GSI, Germany).

Theoretical studies of glueballs are only performed within nonperturbative approaches.
The  bound states of gluons were considered within
the lattice QCD~\cite{Bali:1993fb,Gregory:2012hu,Morningstar:1999rf,Chen:2005mg},
the flux tube model~\cite{Robson:1978iu,Isgur:1984bm},
constituent models~\cite{Jaffe:1975fd,Carlson:1984wq,Chanowitz:1982qj,Cornwall:1982zn,Cho:2015rsa,Boulanger:2008aj},
and in the holographic approach~\cite{Csaki:1998qr,Bellantuono:2015fia,Chen:2015zhh,Brunner:2016ygk}.
The first study~\cite{Novikov:1979ux} of glueballs
in the framework of QCD Sum Rules (SRs)~\cite{Shifman:1978bx} considered a pseudo-scalar $0^{-+}$ state
with an obtained mass of $\sim 1$~GeV.
Later the same group~\cite{Novikov:1979va} applied this
method to a scalar $0^{++}$ glueball state and estimated its mass to be $\sim 0.7$~GeV.
Two-gluon glueballs have been broadly studied using
QCD SRs~\cite{Novikov:1979ux,
 Novikov:1979va,Shuryak:1982dp,Zhang:2003mr,
 Narison:2005wc}.
In further studies~\cite{Harnett:2000fy,Forkel:2003mk}, 
these QCD SRs for the scalar and pseudoscalar glueballs was improved by
calculating the direct instanton contribution and 
the radiative corrections to the perturbative and nonperturbative parts of the correlator.
Three-gluon glueballs were considered in~\cite{Latorre:1987wt} for a $0^{++}$-state
and later in works~\cite{Liu:1998xx,Hao:2005hu} the application of QCD
SRs was extended to the $0^{-+}$ scalar, vector and tensor states.
Further reviews of glueball physics can be found in~\cite{Mathieu:2008me,Ochs:2013gi}.

A way to avoid problems related to the mixing of glueballs with ordinary mesonic states would be
to study glueballs with exotic quantum numbers
($0^{\pm-}$, $1^{-+}$, $2^{+-}$,...)
which are not allowed in quark-antiquark systems.
In our recent study~\cite{Pimikov:2016pag,Pimikov:2017xap} we proposed the $0^{--}$ glueball current of  dimension-12, which was used to obtain 
estimations of the mass, the decay constant and the width of the $0^{--}$ glueball.

In this paper, we present for the first time a detailed procedure for the construction
of the three-gluon glueballs currents based on the helicity formalism
following~\cite{Jacob:1959at,Fritzsch:1975tx,Mandula:1982us,Jaffe:1985qp,Boulanger:2008aj}.
This procedure is applied to construct the $0^{\pm-}$ glueball currents of
the lowest possible dimension.
Using these constructed currents, the QCD SRs have been obtained and analyzed to extract masses
and decay constants of $0^{\pm-}$ glueballs.

The search for the lowest dimension currents
has been motivated by the necessity to improve the reliability of QCD SRs .
In comparison with our previous study~\cite{Pimikov:2016pag},
the SRs presented here have the following improvements:
the Operator Product Expansion (OPE) starts from the condensates of lower dimensions so
that the uncertainties in the OPE can be reduced;
the current of lower dimension leads to a larger coupling with the glueball state;
the first resonance contribution to SRs is larger for the current of lower dimension.
In fact, in the new QCD SR for the $0^{--}$ state, the leading nonperturbative contribution comes
from the 3-gluon condensate $<G^3>$, while in our previous study~\cite{Pimikov:2016pag} the OPE starts from 4-gluon condensates.
From the new SR, we have found that the mass of the $0^{--}$ glueball is very close to our previous result~\cite{Pimikov:2016pag}.
At the same time, the coupling of the new current to the glueball state has been found to be significantly larger compared to the dimension-12 current
suggested in~\cite{Pimikov:2016pag}.
Therefore we conclude that the new current better represents the glueball state.

The paper is organized as follows.
In Sec. II
we present the procedure
for constructing of the three-gluon current
using the helicity formalism~\cite{Jacob:1959at,Fritzsch:1975tx,Mandula:1982us,Jaffe:1985qp,Boulanger:2008aj}.
We construct the currents that couple to the exotic $0^{\pm}-$ glueball helicity states.
In Sec. III we present the OPE of correlators of the new currents and present
the detailed theoretical scheme of QCD SRs.
The masses and decay constant of the $0^{\pm-}$ glueballs are extracted then from QCD SRs.
Section IV contains the discussion of our
results.

\section{Three-gluon currents}

Here we provide the application of the helicity formalism to
the construction of three-gluon currents in general form.
The described technique is applied to construct
the gauge invariant colorless currents that couple to
$0^{\pm-}$ glueball states.

\subsection{Three-gluon helicity states}


The gluon field tensor $G_{\mu\nu}$ corresponds to $(1,0)\oplus(0,1)$
representation of the Lorentz group and can be
decomposed to positive and negative helicity parts $G_{\mu\nu}=G_{\mu\nu}^++G_{\mu\nu}^-$,
where
$
G_{\mu\nu}^{\mp}=(G_{\mu\nu} \pm \tilde G_{\mu\nu})/2
$
and dual tensor
${\tilde G_{\mu\nu}=-i\epsilon_{\mu\nu\alpha\beta}G^{\alpha\beta}/2}$.
The negative helicity strength tensor $G^-$ is in the $(1, 0)$ representation,
 and the positive-helicity strength tensor $G^+$ is in the $(0, 1)$ representation,
 thus the different helicity tensors are not mixed under Lorentz transformations.
Therefore using helicity strength tensor $G_{\mu\nu}^\pm$
as building blocks allows to decompose the glueball currents
into irreducible representations of the Lorentz group~\cite{Jaffe:1985qp}.

To consider the three gluon helicity current in a general form,
we define the generating current as: 
\begin{eqnarray}\label{eq:helicity-current}
&&J(G_1G_2G_3) \sim  \\\nn
&& \frac 1{3!} S_{G_1G_2G_3}g_s^3
(O_{1} G_{1\mu_1\nu_1})^{a_1}
(O_{2} G_{2\mu_2\nu_2})^{a_2}
(O_{3} G_{3\mu_3\nu_3})^{a_3}\,,
\end{eqnarray}
where $G_i$ with $i=1,2,3$ stands for the gluon field strength tensor in one of
the following forms:
the strength tensor $G$,
the dual tensor $\tilde G$,
the positive helicity tensor  $G^+$ or
the negative helicity tensor  $G^-$.
The operator of symmetrization $S_{G_1G_2G_3}$ ensures that
the current is symmetrical with respect to gluon interchange.
The operators $O_i$ with $i=1,2,3$ are the product of covariant derivatives
to respect the gauge invariance of the constructed currents:
\begin{eqnarray}\label{eq:operatorO}
O_i G_{\mu\nu} = D_{\tau_{1}}D_{\tau_{2}}\cdots D_{\tau_{n}} G_{\mu\nu}\,.
\end{eqnarray}
 In order to consider both C-parities,
we omit here the trace ${\rm Tr}$ in the color space
that will be recovered later to construct colorless currents
and insure the gauge invariance.
Taking various $O_i$ and ways for the contraction of the Lorentz indices,
the currents of various quantum numbers will be generated.

There are two possible combinations to construct helicity-$\lambda$ current $J^P_\lambda$ of the  parity-$P$
that are symmetrical with respect to the gluon exchanges:
the maximal helicity ($\lambda =3$) state with the parities $P=\pm 1$:
\begin{eqnarray}\nn
J^{\pm}_3=J_{+++}\pm J_{---}\,,
\end{eqnarray}
and the minimal helicity ($\lambda =1$) current with the  parities $P=\pm 1$:
\begin{eqnarray}\nn
J^{\pm}_1=J_{++-}\pm J_{--+}\,,
\end{eqnarray}
where the indices in the currents on the right-hand side of the equation
mean the helicities of gluons as
$J_{h_1h_2h_3}=J(G^{h_1}G^{h_2}G^{h_3})$
in the general form, see Eq.\eqref{eq:helicity-current}.
In the definitions of the maximal and minimal helicity current we have
omitted for simplicity the sign $C$ of the arbitrary charge parity $J_\lambda^{P}=J_\lambda^{PC}$.
Expanding the helicity currents in terms of the gluon strength tensor and
its dual tensor one finds:
\begin{eqnarray}\nn
J^{+}_3&=&\frac 14\left(J(GGG)+J(G\tilde G\tilde G)+J(\tilde GG\tilde G)+J(\tilde G\tilde GG)\right)\,;\\\nn
J^-_3&=&-\frac 14
 \left(
  J(\tilde G\tilde G\tilde G)+J(\tilde GGG)+J(G\tilde GG)+J(GG\tilde G)
 \right)\,;\\\nn
J^+_1&=&\frac 1{12}\left(3J(GGG)-J(G \tilde G\tilde G)-J(\tilde GG\tilde G)-J(\tilde G\tilde GG)\right)\,;\\\nn
J^-_1&=&\frac 1{12}
\left(3J(\tilde G\tilde G\tilde G)-J(\tilde GGG)-J(G\tilde GG)-J(GG\tilde G)
\right)\,.
\end{eqnarray}
In this consideration the three-gluon $0^{\pm+}$ glueball currents~\cite{Latorre:1987wt,Hao:2005hu}:
\begin{eqnarray}\label{eq:0pp}
J^{++}&=&g_s^3 f^{abc}G^a_{\mu\nu}G^b_{\nu\rho}G^c_{\rho\mu}\,,\\\label{eq:0mp}
J^{-+}&=&g_s^3 f^{abc}\tilde G^a_{\mu\nu}\tilde G^b_{\nu\rho}\tilde G^c_{\rho\mu}
\end{eqnarray}
represent the maximal ($\lambda=3$) helicity states $J^\pm_3=J^{\pm+}$
while all minimal ($\lambda=1$) helicity states have $J^\pm_1=0$.
By introducing arbitrary linear operators $O_i$
these currents, Eqs. (\ref{eq:0pp}) and (\ref{eq:0mp}),
can be generalized in following form
\begin{equation}\label{eq:current:gen2}
J(G_1G_2G_3) \sim g_s^3
(O_{1} G_{1\mu\nu})^{a_1}
(O_{2} G_{2\nu\rho})^{a_2}
(O_{3} G_{3\rho\mu})^{a_3}\,.
\end{equation}
This form of the current has been used in
the first QCD SR based study of negative charge parity $0^{--}$
scalar glueballs~\cite{Pimikov:2016pag}.
One can see that the contraction of the Lorentz indices leads to the
following property for this type of currents, Eq. (\ref{eq:current:gen2}):
\begin{eqnarray}\nn 
J(GGG)=J(G\tilde G\tilde G)=J(\tilde GG\tilde G)=J(\tilde G\tilde GG)\,;\\\nn
J(\tilde G\tilde G\tilde G)=J(\tilde GGG)=J(G\tilde GG)=J(GG\tilde G)\,.
\end{eqnarray}
Therefore such a currents represents maximal helicity states.

\subsection{Three-gluon helicity states of $0^{\pm-}$  glueballs}
In order to construct the gauge invariant currents that couple to $0^{\pm-}$ glueballs,
we are looking for scalar or unconserved vector local currents.
The conserved vector currents correspond to the spurious state and do not couple
to the scalar state~\cite{Jaffe:1975fd}.
Another important requirement to the current
is having the nonzero Leading Order (LO) perturbative contribution to the spin-0 part of the correlator.
In configuration space, the spin-0 projector in the correlator is a partial derivative.
Therefore, the conserved vector currents have no spin-0 contribution.
To eliminate possible ambiguity in the construction of the current
and to avoid spurious states,
we consider only the currents that are defined by the helicity gluons
field strength tensor adopting the helicity formalism~\cite{Jacob:1959at,Fritzsch:1975tx,Mandula:1982us,Jaffe:1985qp,Boulanger:2008aj}.
To construct the lowest dimension
currents from helicity gluons that couple to $0^{\pm -}$ glueball states,
we propose the generating current that respects all requirements described above:
\begin{eqnarray}\label{eq:GGGdim9}
&&J_\alpha(G_1G_2G_3) =
\\\nn &&~~~~~~
\frac {2g_s^3}{3!}S_{123}  \Tr\left(
\{(D_{\rho} G_{1\mu\nu}),
(D_{\sigma} G_{2\rho\nu})
\}
(D_{\mu} G_{3\sigma\alpha})
\right)\,,
\end{eqnarray}
where the factor $2g_s^3$ was introduced to have at LO
\begin{eqnarray}\nn
J_\alpha(GGG)  \stackrel{\text{LO}}{=} g_s^3 d^{abc}
(\partial_{\rho} G^a_{\mu\nu})
(\partial_{\sigma} G^b_{\rho\nu})
(\partial_{\mu} G^c_{\sigma\alpha})
\,,
\end{eqnarray}
that can be easily compared to the currents, Eq.~(\ref{eq:0pp}) and Eq.~(\ref{eq:0mp}),
suggested in~\cite{Latorre:1987wt,Hao:2005hu} for $0^{\pm+}$ glueball states.

The currents Eq.(\ref{eq:GGGdim9}) of the maximal ($\lambda=3$) helicity state
appear to be conserved in LO $\partial_\alpha J^{\pm-}_{3,\alpha}=0$
and therefore the maximal helicity current does not respect
the nonzero LO term condition.
While the minimal ($\lambda=1$) helicity $0^{\pm -}$ currents
based on the generating current Eq.(\ref{eq:GGGdim9})
are non-conserved currents and have all desired properties:
\begin{eqnarray}\label{eq:GGGdim9heli}
J^{+-}_{\alpha}=
g_s^3 \Tr (
\{
(D_{\tau} G_{\mu\nu}),
(D_{\tau} G_{\rho\nu})\}
(D_{\mu} G_{\rho\alpha})
)\,,\\\nn
J^{--}_{\alpha}=
g_s^3 \Tr (
\{
(D_{\tau} G_{\mu\nu}),
(D_{\tau} G_{\rho\nu})\}
(D_{\mu} \tilde G_{\rho\alpha})
)\,.
\end{eqnarray}
We propose these currents to study $0^{\pm -}$ states.
The new current for the $0^{--}$ state has significantly lower dimension than
the current suggested recently in~\cite{Pimikov:2016pag}.
As we see in the next section and discuss in the introduction,
the reduction of the dimension leads to the improvements of the reliability of QCD SRs:
it reduces the OPE uncertainties and increases
the coupling with the state and
the first resonance contribution to SR.

Any other choice of the dimension-9 generating current Eq.~(\ref{eq:GGGdim9}),
leads to the zero current or to the alternative current $J^{\pm-}_{\alpha,\text{alt}}$
that has identical coupling to spin-0 state in LO:
$
\partial_\alpha J^{\pm-}_{\alpha,\text{alt}}
\stackrel{\text{LO}}{=}
\partial_\alpha J^{\pm-}_{\alpha}$.
Using the gluon field tensors and covariant derivatives to ensure a gauge invariance of the current,
we did not find any three-gluon current of dimension-7 and dimension-8
that respect above requirement.
 Applying the helicity formalism to the four-gluon states leads us to the conclusion
 that there is no any nonzero helicity current of dimension-8
 which couples to the exotic $0^{\pm -}$ glueballs.

\section{Sum Rules}

\subsection{OPE of correlators}

Here we present the result for OPE of correlators that is
the theoretical basis of QCD SRs approach~\cite{Shifman:1978bx}:
\begin{equation}\nn 
\Pi^{\pm}_{\mu\nu}(q) = i\int\!\! d^4x\, e^{iqx} \va{J^{\pm-}_{\mu}(x) J^{\pm-}_{\nu}(0)^\dagger}\,
\end{equation}
where the proposed current $J^{\pm-}_{\alpha}$ is given by Eq.~(\ref{eq:GGGdim9heli}) and
couples to the gluonic bound state $|G(0^{\pm -})\rangle$
with the mass $m_{\pm}$  and the decay constant $f_{\pm}$ through the relation:
\begin{eqnarray}\nn 
\va{0|J^{\pm-}_{\alpha}|G(0^{\pm -})}=p_\alpha f_{\pm} m_{\pm}^{6}\,.
\end{eqnarray}
The correlators of the vector currents have two components:
\begin{eqnarray}\nn
\Pi^{\pm}_{\mu\nu}(q)=\Pi^{(1)\pm}(q^2)(q_\mu q_\nu-q^2g_{\mu\nu})+\Pi^{(0)\pm}(q^2)q_\mu q_\nu\,,
\end{eqnarray}
where $\Pi^{(0)}$ and $\Pi^{(1)}$ are spin-0 and spin-1 contributions, respectively.
\begin{figure*}[t]
 \centerline{
  \includegraphics[width=0.18\textwidth]{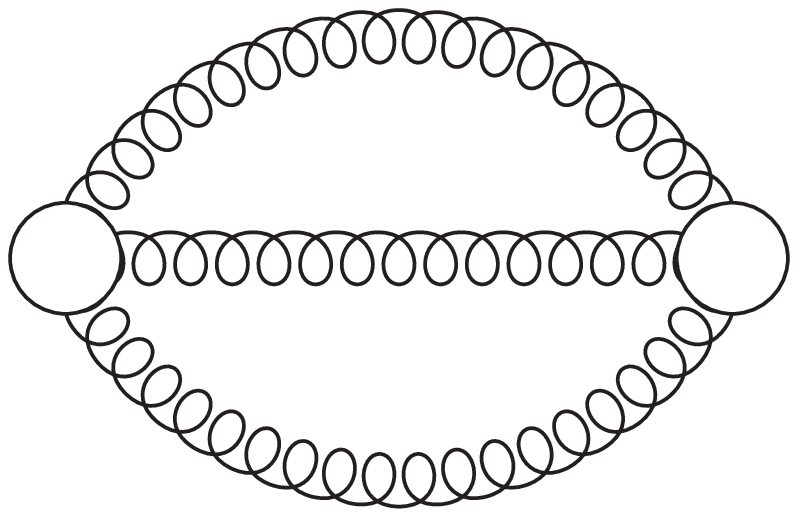}~
  \includegraphics[width=0.18\textwidth]{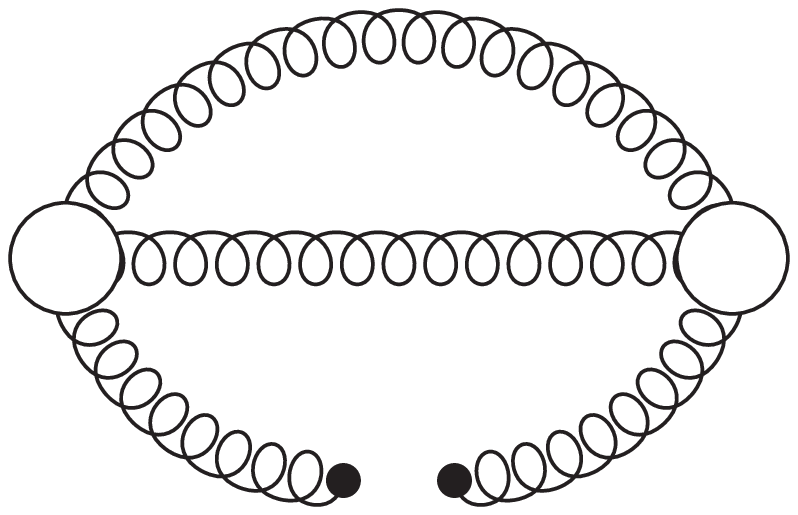}~
  \includegraphics[width=0.18\textwidth]{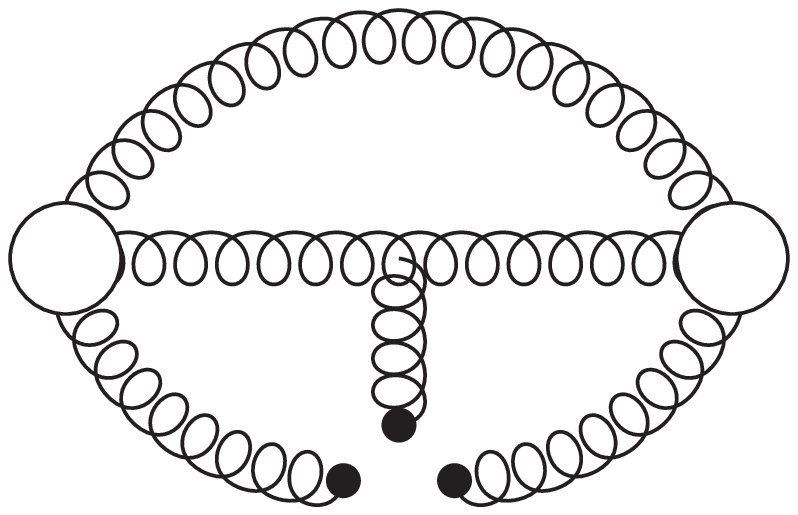}~
  \includegraphics[width=0.18\textwidth]{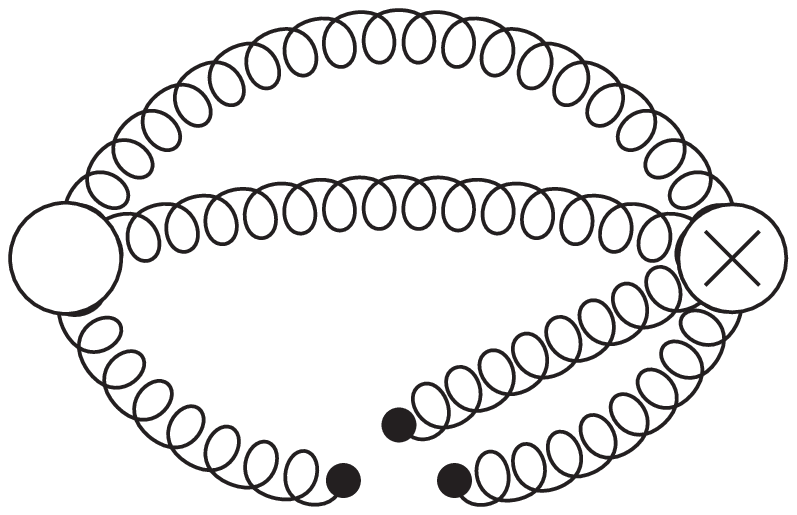}~
  \includegraphics[width=0.18\textwidth]{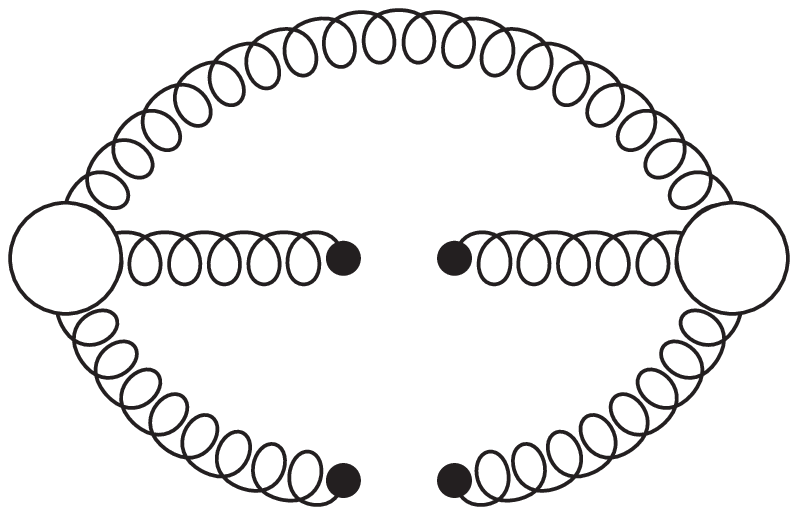}
 }
 \caption{
  \label{fig:diagr}
  Generic diagrams for the contributions to OPE of the correlator $\Pi^{\pm}_{\mu\nu}(q)$.
  The first diagram represents the LO perturbative term,
  while the remaining part of diagrams depicts the nonperturbative contributions.
  For simplicity, we use the nonlocal condensate notation~\cite{Mikhailov:1986be,Mikhailov:1991pt,Grozin:1985wj,Grozin:1994hd} for graphical
  representation of various contributions coming from standard (local) condensates.
  The black dots represent the nonlocal condensate that gives the contribution to the higher dimension condensates.
  Three nonperturbative diagrams accumulate all possible contribution related to
  the Taylor expansion of condensate gluon fields (denoted by the black dots).
  The fourth diagram corresponds to the terms with an extra vacuum gluon coming from any of the covariant derivatives in the current.
  The last diagram term starts from dimension-12 condensates
   and therefore has been omitted.
 }
\end{figure*}

Here we consider only the spin-0 part of the correlator OPE up to dimension-8 condensates:
\begin{eqnarray}\nn
\Pi^{(0)\pm}_\text{(OPE)}=
\Pi^{(0)\pm}_\text{(pert)}
+\Pi^{(0)\pm}_\text{(G2)}
+\Pi^{(0)\pm}_\text{(G3)}
+\Pi^{(0)\pm}_\text{(G4)}+\cdots\,,
\end{eqnarray}
where the following terms are considered: the LO perturbative term (pert),
the dimension-4 (G2), the dimension-6 (G3), and dimension-8 (G4) nonperturbative terms.
The terms of the correlator OPE have been calculated and are given as follows:
\begin{eqnarray}\nn
\Pi^{(0)\pm}_{\text{(pert)}}  &=&
\frac{-5\alpha_s^3}{9!8\pi}q^{12}\ln\frac{-q^2}{\mu^2}\,,~~~
\Pi^{(0)\pm}_\text{(G2)}  = 0 \,,\\\label{eq:Pi0mm:dim9helicit}
\Pi^{(0)\pm}_\text{(G3)}  &=&
\pm\frac{5\alpha_s^2}{2^8 3^3}G_{G3}^{\pm} \cdot
q^{6}\ln\frac{-q^2}{\mu^2}
\,,\\\nn
\Pi^{(0)\pm}_\text{(G4)}  &=&
\mp\frac{\alpha_s^2 \pi^2}{2^6 3^3}\va{\alpha_s^2 G^4}_\pm\cdot
q^{4}\ln\frac{-q^2}{\mu^2}
\,,
\end{eqnarray}
where $\alpha_s=g_s^2/(4\pi)$ is the coupling constant, $\mu$
is the renormalization scale.
The contributions $G_{G3}^{\pm}$ and $\va{\alpha_s^2 G^4}_\pm$
are linear combinations of the dimension-6 and dimension-8 condensates described below.
We adopt Mathematica package FEYNCALC~\cite{Shtabovenko:2016sxi} to handle
the algebraic manipulation.
The LO perturbative term is represented by the two-loop sunset diagram (the first diagram in Fig. \ref{fig:diagr}),
therefore for any scalar three-gluon current the largest prime divisor of denominator must be less than the dimension of the current.
The leading nonperturbative contribution
from the nonlocal two gluon condensate~\cite{Mikhailov:1986be,Mikhailov:1991pt,Grozin:1985wj,Grozin:1994hd}, represented by second diagram in
Fig. \ref{fig:diagr}, is defined by the dimension-6 local condensates thanks to the derivatives in the currents:
\begin{eqnarray}\nn
&&G_{G3}^{+}=9\va{g^3G^3}-88\pi \alpha_s \va{J^2} \,, 
\\\nn
&&G_{G3}^{-}=9\va{g^3G^3}-20\pi \alpha_s \va{J^2}\,,
\end{eqnarray}
where notations for condensates of dimension-6 are $\va{g^3G^3}=\va{g^3f^{abc}G^a_{\mu\nu}G^b_{\nu\ro}G^c_{\ro\mu}}$ and
$\va{J^2}=\va{J^a_\mu J^a_\mu}$ with the quark current
$J^a_\mu=\bar q\ga_\mu t^aq$.
For the same reason, the leading term of the third and fourth diagrams
shown in Fig. \ref{fig:diagr} is dimension-8 contribution.
While the last diagram in Fig. \ref{fig:diagr} starts from dimension-12 condensate and therefore is not considered here.
The four-quark condensate $\va{J^2}$ is considered to be insignificant
compare to three-gluon condensate $\va{gG^3}\gg \va{J^2}$
and has not been included in the QCD SRs analysis.
Therefore, the quarks contribute only perturbatively due to the strong coupling evolution as
it is discussed below (see Eq. (\ref{eq:constants})).
The total dimension-8 condensate contribution to the correlator are presented by
the four-gluon condensates:
\begin{eqnarray}\nn
&&\va{\alpha_s^2G^4}_+=
\\\nn &&~~~~
155\va{(\alpha_s f^{abc}G^b_{\mu\nu}G^c_{\rho\sigma})^2}
+2678\va{(\alpha_s f^{abc}G^b_{\mu\nu}G^c_{\nu\rho})^2}\,,\\\nn
&&\va{\alpha_s^2 G^4}_-=
\\\nn &&~~~~
845\va{(\alpha_s f^{abc}G^b_{\mu\nu}G^c_{\rho\sigma})^2}
+1298\va{(\alpha_s f^{abc}G^b_{\mu\nu}G^c_{\nu\rho})^2}\,,
\end{eqnarray}
where quark-gluon condensates have been omitted.
As expected, the nonperturbative terms in
the approximation of self-dual (SD) gluon fields are equal in absolute value
and have different signs (see Eq. (\ref{eq:Pi0mm:dim9helicit})) for the parity $P=\pm 1$:
\begin{eqnarray}\nn
&&G_{G3}^{\pm} \stackrel{\text{SD}}{~=~} 9\va{g^3f^{abc}G^a_{\mu\nu}G^b_{\nu\ro}G^c_{\ro\mu}}\,, 
\\\nn
&&\va{\alpha_s^2 G^4}_\pm\stackrel{\text{SD}}{~=~}
2^2 3^2 83 \va{(\alpha_s f^{abc}G^b_{\mu\nu}G^c_{\nu\rho})^2}\,.
\end{eqnarray}
For QCD SRs analysis we apply  the hypothesis of vacuum dominance (HVD)
to estimate the dimension-8 condensate:
\begin{eqnarray}\label{eq:HVDresult}
&&\va{\alpha_s^2 G^4}_+\stackrel{\text{HVD}}{~=~} k_\text{HVD}\frac{3}{2^4}1151 \va{\alpha_s G^2}^2    \,,~~~
\\\nn
&&\va{\alpha_s^2 G^4}_-\stackrel{\text{HVD}}{~=~} k_\text{HVD}\frac{3\cdot7}{2^4}263 \va{\alpha_s G^2}^2    \,,
\end{eqnarray}
where $k_\text{HVD}$ denotes the coefficient of the HVD factorization violation.
We vary this coefficient in the range $k_\text{HVD}\in[0.25,4]$ to include
the HVD-related uncertainty.
Evaluating QCD SRs, we apply the results of recent studies \cite{Narison:2011xe,Narison:2011rn}
where the charmonium moments sum rules has been used to obtain
the gluon condensate estimations:
\begin{eqnarray}\label{eq:condensateG2G3}
&&\va{g^3G^3}=(8.2\pm 2.0)~\text{GeV}^2 \va{\alpha_s G^2}\,, 
\\\nn
&&\va{\alpha_sG^2}=0.07(2) ~\text{GeV}^4\,.
\end{eqnarray}
The ratio between the three-gluon and the two-gluon condensates agrees well with
the instanton model~\cite{Schafer:1996wv}
for the instanton radius  $\rho_c=1/(600~\text{MeV})$:
\begin{eqnarray}\nn 
\va{g^3G^3}=
\frac{48 \pi}{5\rho_c^2}\va{\alpha_s G^2}\,.
\end{eqnarray}
Due to the large value of the Borel parameter $M^2$ in QCD SRs (see bellow) 
for exotic glueballs the possible direct instanton contributions
to the correlators are expected to be strongly suppressed
in comparison to OPE terms and therefore are not considered here.

\begin{figure*}[bt]
 \includegraphics[height=0.3\textwidth]{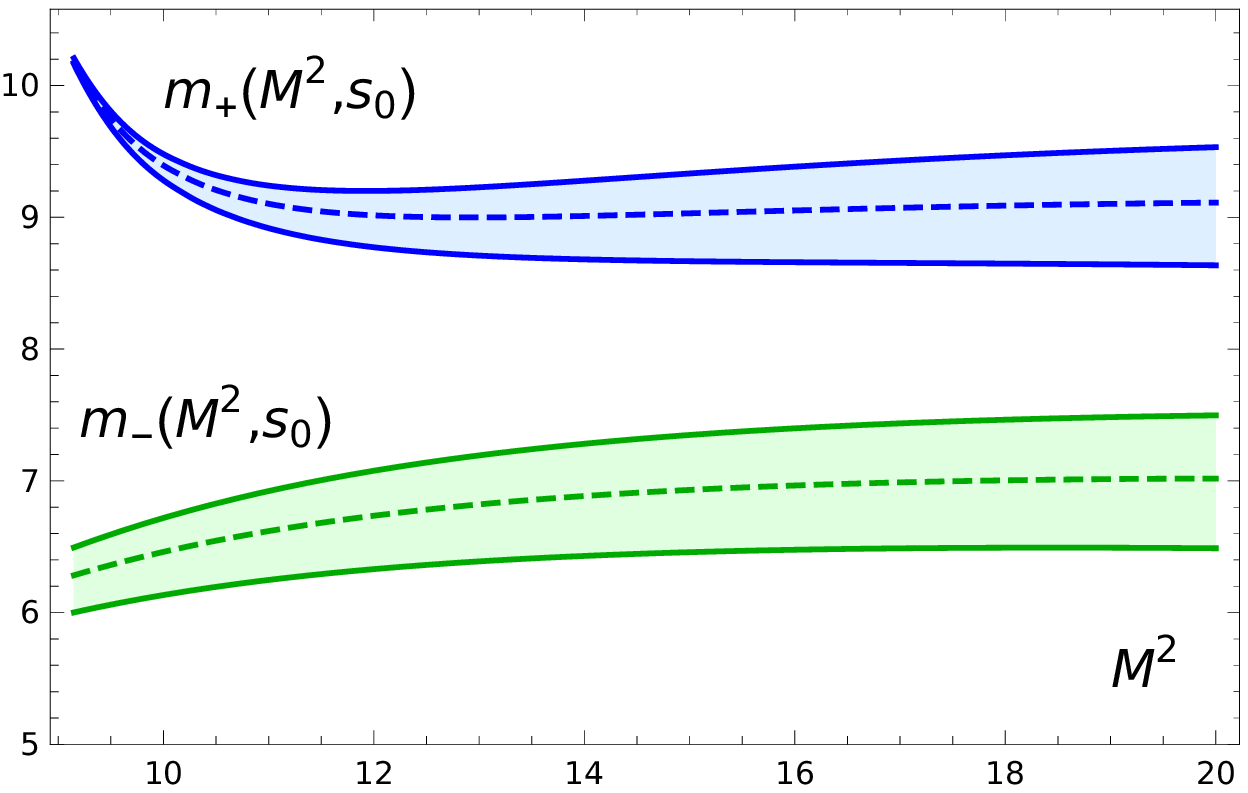}
 \hfill
 \includegraphics[height=0.3\textwidth]{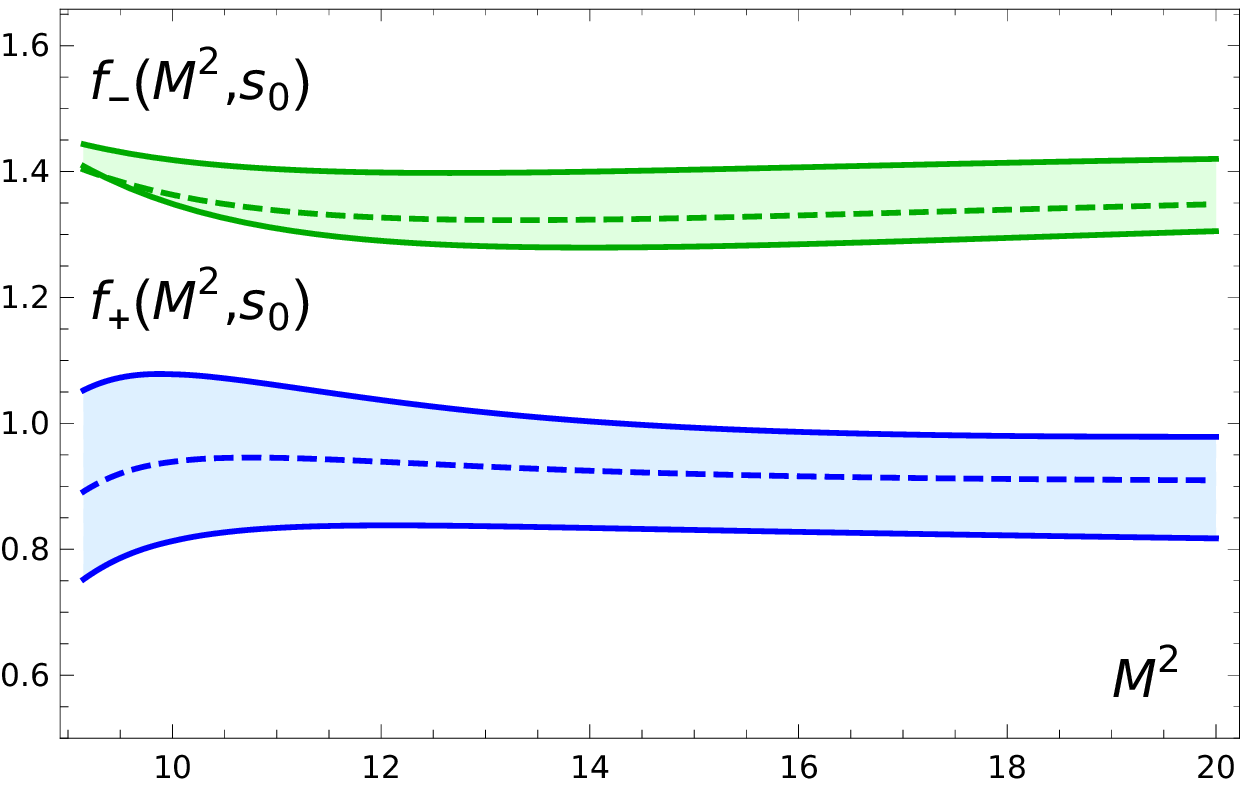}
 \caption{
  \label{fig:M2depend}
  The Borel parameter dependence of the mass (left panel) and the decay constant (right panel)
  with the central values of the gluon condensates presented for
  the scalar state by the blue dashed band and
  for the pseudoscalar state by the green dashed band.
  The dashed bands represent the threshold variation of the Borel parameter dependence.
  The thresholds vary in the fiducial intervals presented in Table~\ref{tab:SRresults}.
 }
\end{figure*}

\subsection{QCD SRs}
We analyze the constructed QCD SRs for the $0^{\pm-}$ states on the same footing.
Therefore here and below for simplicity we omit the parity and the spin signs 
$\Pi^{(0)\pm}_{t} \to \Pi_{t}$,
where $t$ denotes the different contributions to OPE  of the correlator
as explained above Eqs.(\ref{eq:Pi0mm:dim9helicit}).
In simplified notation the truncated OPE of correlator has a form:
\begin{eqnarray}\nn
\Pi_\text{(OPE)}= \Pi_\text{(pert)}
+\Pi_\text{(G3)}+\Pi_\text{(G4)}\,.
\end{eqnarray}
The phenomenological part of QCD SR is based on the modeling of spectral density.
For the phenomenological description of the correlator,
we use the one-resonance model with the continuum contribution modeled by Im-part of 
the correlator OPE:
\begin{eqnarray}\nn 
&&\text{Im} \Pi_{\text{(ph)}}(-s)
=\\\nn&& ~~~~~
\pi m^{12}f^2 \delta(s-m^2)
+\Theta(s-s_0)\text{Im} \Pi_{\text{(OPE)}}(-s)\,,
\end{eqnarray}
where 
$m$ is the mass of a resonance and $s_0$ is the continuum threshold.
Then QCD SR reads
\begin{eqnarray}\label{SR1}
\frac{1}{\pi}\int_0^{s_0}\frac{\text{Im}\Pi_{\text{(OPE)}}(-s)}{s+Q^2}ds
&=&
\frac{f^2m^{12}}{m^2+Q^2}\,.
\end{eqnarray}
In the framework of QCD SRs~\cite{Shifman:1978bx},
the Borel transform~$\hat{B}$
\begin{eqnarray}\nn
\hat{B}_{Q^2\to M^2}\!\left[\Pi(Q^2)\right]
= \mathop{\text{lim}}\limits_{n\to\infty}\!
\frac{(-Q^2)^n}{\Gamma(n)}\!
\left[\frac{d^n}{dQ^{2n}}\Pi(Q^2)\right]_{Q^2=n M^2}\,,
\end{eqnarray}
is applied to both sides of the SR, Eq. (\ref{SR1})
in order to reduce the SR uncertainties by suppressing the contributions
from excited resonances and higher order OPE terms.
The Borel transformation modifies the components of the sum rule:
\begin{eqnarray}\nn 
{\cal R}^t_0(M^2,s_0)&=&
\frac 1\pi \int_0^{s_0}\!\!ds~ \text{Im} \Pi_t(-s)~e^{-s/M^2}\,,\\\nn
{\cal R}^\text{(res)}_0(M^2)&=&
m^{12} f^2 e^{-m^2/M^2}\,. 
\end{eqnarray}
Here we follow common practice of renormalization group improvement after Borel transformation,
therefore in $\text{Im} \Pi^t(-s)$ all coupling constant are replaced by running constant $\alpha_s\to\alpha_s(M^2)$:
\begin{eqnarray}\label{eq:constants}
\alpha_s(Q^2)=\frac{4\pi}{b_0 \ln(Q^2/\Lambda_\text{QCD}^2)}\,,~
\end{eqnarray}
where the beta-function LO coefficient $b_0=11-2N_f/3$,
the QCD scale $\Lambda_\text{QCD}=350~\text{MeV}$, and
number of the flavors $N_f=4$.

The mass is extracted from the family of the derivative SRs defined by
\begin{eqnarray}\nn 
{\cal R}^t_k(M^2,s_0)&=& M^4\frac{d}{d M^2}{\cal R}^t_{k-1}(M^2,s_0) \,.
\end{eqnarray}
Denoting by ${\cal R}^\text{(SR)}$ the difference of the OPE result
and the continuum contribution for any $k\geq 0$:
\begin{eqnarray}\nn 
&&{\cal R}^\text{(SR)}_k(M^2,s_0) =\\\nn
&& ~~~~~
{\cal R}^\text{(pert)}_k(M^2,s_0)
+{\cal R}^\text{(G3)}_k(M^2,s_0)
+{\cal R}^\text{(G4)}_k(M^2,s_0) \,.
\end{eqnarray}
we define the master sum rule  ($k=0$) and the derivative SRs ($k>0$) by
the following equations:
\begin{eqnarray}\label{eq:RESvsSR}
{\cal R}^\text{(SR)}_k(M^2,s_0) &\approx&  {\cal R}^\text{(res)}_k(M^2,s_0)\,.
\end{eqnarray}

The high dimension of the considered currents leads to
the dependency of the continuum spectral density on $s$ as
${\text{Im} \Pi_{\text{(OPE)}}\sim s^6}$.
Therefore, having in mind that the continuum contribution could
give the large contribution~\cite{Matheus:2006xi,Huang:2016rro,Palameta:2017ols},
we define the upper boundary $M^2<M_+^2(s_0)$ of the fiducial window
by the following condition which is less restrictive than
the condition for the low-dimension correlators
suggested in~\cite{Shifman:1978bx}:
\begin{eqnarray}\label{eq:window.R}
\frac{{\cal R}^\text{(res)}_k(M^2)}{{\cal R}^\text{(SR)}_k(M^2,\infty)}
&\approx&
\frac{{\cal R}^\text{(SR)} _k(M^2,s_0)}{{\cal R}^\text{(SR)}_k(M^2,\infty)}
>  \frac 1{10}\,.
\end{eqnarray}
This condition influences the definition of the SR uncertainty,
while the central values of predictions appear to be insensitive to it.
The lower  boundary $M_-^2$ of the fiducial window $M^2\in [M_-^2,M_+^2(s_0)]$
is limited by the conditions
\begin{eqnarray}\label{eq:window.L}
 \frac{|{\cal R}^\text{(G3)}_k(M^2,\infty)|}{{\cal R}^\text{(pert)}_k(M^2,\infty)}
 &<&  \frac 23\,, ~
 \frac{|{\cal R}^\text{(G4)}_k(M^2,\infty)|}{{\cal R}^\text{(pert)}_k(M^2,\infty)}
 <  \frac 13\,,
\end{eqnarray}
that insure the OPE reliability.

The values of mass and the decay constant can be extracted from QCD SRs,
 Eq.(\ref{eq:RESvsSR}), as:
\begin{eqnarray}\nn 
m_k(M^2,s_0) &=&\sqrt{\frac{{\cal R}^\text{(SR)}_{k+1}(M^2,s_0)}{{\cal R}^\text{(SR)}_{k}(M^2,s_0)}}\,,\\\nonumber
f_k^2(M^2,s_0) &=&
\frac{e^{M_G^2/M^2}{\cal R}^\text{(SR)}_k(M^2,s_0)}{M_G^{2(6+k)}}\,.
\end{eqnarray}

\begin{table*}[ht]
 \caption{\label{tab:SRresults}
  QCD SRs results on the masses, the decay constants of $0^{\pm-}$ glueballs
  for the $k=0$ case
  are presented together with corresponding fiducial intervals of the SR parameters:
  the Borel parameter $M^2$ (the upper limit is given for the central value of the threshold)
  and the threshold value $s_0$.
  The fiducial intervals have been defined through the conditions Eqs.(\ref{eq:window.R},\ref{eq:window.L},\ref{eq:window.s0}).
   Three given uncertainties of the mass and the decay constant values are estimated by
   the Borel parameter $M^2$ variation,
   the threshold $s_0$ dependence
   and the gluon condensates uncertainties.
 }{
  \begin{ruledtabular}
   \begin{tabular}{ccccc}
    state &  mass, GeV   & decay constant, MeV & $M^2$, GeV$^2$ & $s_0$, GeV$^2$
    \\\hline
    $0^{--}$  &
    $6.84\pm 0.36~^{+0.44}_{-0.47}~^{+0.26}_{-0.37}$  &
    $1.34\pm 0.04~^{+0.07}_{-0.03}\pm 0.02$  & $[9.2, 21.5]$ & $78\pm 10$
    \\\hline
    $0^{+-}$  &
    $9.23\pm 0.56 \pm 0.38 ^{+0.40}_{-0.47}$  &
    $0.93\pm 0.02~^{+0.08}_{-0.10}\pm 0.02$  & $[9.2, 30.0]$ & $120\pm 14$
   \end{tabular}
 \end{ruledtabular}}
\end{table*}

We define the mass and the decay constant
 by keeping the $M^2$-stability criteria $\delta_k$ below $10\% \sim 1/3^2$
that is the assumed OPE accuracy related to the condition Eq. (\ref{eq:window.L}):
\begin{eqnarray}\label{eq:window.s0}
\delta_k=
\frac{\text{max} f_k^2(M^2,s_0) -\text{min} f_k^2(M^2,s_0)}
     {\text{max} f_k^2(M^2,s_0) +\text{min} f_k^2(M^2,s_0)}
     <1/10\,.
\end{eqnarray}
This condition puts limits on the continuum threshold value $s_0$.
The conditions Eqs.(\ref{eq:window.R},\ref{eq:window.L},\ref{eq:window.s0})
define the fiducial set of $(M^2,s_0)$-values.
Finally  we define the prediction for the mass and the decay constant as
an average of the maximal and the minimal values on the fiducial interval of $M^2$
with the fixed central value of threshold given in the last column of Table~\ref{tab:SRresults}:
\begin{eqnarray}\nn 
m_k  &=&\frac{\text{max}~m_k(M^2,s_0) + \text{min}~m_k(M^2,s_0)}{2}\,,\\\nn
f_k^2&=&\frac{\text{max}~f_k^2(M^2,s_0) + \text{min}~f_k^2(M^2,s_0)}{2}\,.
\end{eqnarray}
The variation of the mass and the decay constant 
in the fiducial $(M^2,s_0)$-set defines uncertainties coming from
the OPE truncation and the spectral function modeling.

\subsection{QCD SRs results for the $0^{\pm-}$ glueball states}

Performing the QCD SRs analysis described above, we obtain
predictions for the masses and decay constants of the $0^{\pm-}$-states.
These are presented in Table~\ref{tab:SRresults} for the $k=0$ case together
with the fiducial intervals of the SR parameters: the Borel parameter $M^2$
and the threshold value $s_0$.

There are three sources of errors for mass and decay constant presented in  Table~\ref{tab:SRresults}:
the first error represents the  SR stability triggering Borel parameter $M^2$ dependence,
the second represents the threshold $s_0$ dependence
and the third  is the uncertainty related to
the variations of the gluon condensates $\va{G^3}$ and $\va{G^4}$.
The first two errors, which originate from OPE truncation and
continuum modeling,
are defined by variation of results
on the fiducial $(M^2,s_0)$-set that represents the conditions
Eqs.~(\ref{eq:window.R},\ref{eq:window.L},\ref{eq:window.s0}).
The variation of the $\va{G^3}$ condensate comes
from~\cite{Narison:2011xe,Narison:2011rn} (see Eq.~(\ref{eq:condensateG2G3})).
The uncertainties of the $\va{G^4}$ contribution have been estimated from
the variation of the HVD violation coefficient
(see Eq.~(\ref{eq:HVDresult})) and the variation of the two-gluon condensate
$\va{G^2}$ was estimated in~\cite{Narison:2011xe,Narison:2011rn}.
In Fig.~\ref{fig:M2depend}, we present the $k=0$ results for the glueball mass and
the decay constant as a function of the Borel parameter for various values of
the threshold parameter.
As one can see, there is a rather good stability plateau for both quantities which
is ensured by the condition in Eq.~(\ref{eq:window.s0}).
The masses and decay constants estimated with the higher values of $k=1,2,3$ are in
agreement with the $k=0$ case.

\section{Summary}
We have performed a study of C-odd scalar and pseudoscalar exotic glueball states
within the framework of QCD SRs.
The constructed QCD SRs include LO perturbative term and the nonperturbative contributions up to dimension-8 gluon condensates.
The results from the QCD SRs analysis on the masses $m_\pm$ and decay constants $f_\pm$
of the $0^{\pm-}$ glueballs are given as follows: for the pseudoscalar state
\begin{eqnarray} \label{eq:finalRes}
m_{-}=6.8^{+1.1}_{-1.2}~\text{GeV}\,,  f_{-}=1.3\pm 0.1 ~\text{MeV}\,,
\end{eqnarray}
and for the scalar state
\begin{eqnarray}\nn
m_{+}=9.2^{+1.3}_{-1.4}~\text{GeV}\,,
f_{+}=0.9\pm  0.1~\text{MeV}\,.
\end{eqnarray}
The construction of the three-gluon currents has been addressed in general form on the basis of the helicity formalism.
The developed techniques of the helicity based current construction have been used to
build new three-gluon currents of minimal dimension that couple to the $0^{\pm-}$ glueball states.

Our previous QCD SRs results~\cite{Pimikov:2016pag} on the $0^{--}$-glueball mass
using a dimension-12 current, $M_G=6.3^{+0.8}_{-1.1}$ GeV,
is in good agreement with our new estimation, Eq.(\ref{eq:finalRes}).
As one would expect, the current with higher dimension
leads to a smaller coupling to the dimension-12 current with the glueball state~\cite{Pimikov:2016pag}:
$F_G=67 \pm 6$ keV.
Therefore, the new current of minimal dimension represents the most possible configuration of $0^{--}$ glueball.

The Belle Collaboration \cite{Jia:2016cgl} has performed a search in
the range of masses lower than our predicted mass and
found no evidence for the exotic $0^{--}$ glueball.
Our result on the mass of the exotic $0^{--}$ glueball is in 
qualitative agreement with the result of lattice QCD~\cite{Gregory:2012hu}.
 On the other hand, the obtained mass of the $0^{+-}$ glueball state is noticeably larger than the lattice
 results~\cite{Morningstar:1999rf,Chen:2005mg}.
 Unfortunately, the status of exotic glueball masses calculated using lattice QCD not clear at the present time (see the discussion
 in \cite{Gregory:2012hu}  and Table 3 therein). 
 Some lattice groups have seen exotic glueball signals, 
 while others found no indication of any signals
 for the same exotic states.
 Furthermore, in \cite{Gregory:2012hu} it was emphasized  that lattice QCD calculations using heavy glueball degrees of freedom should use improved
 techniques to assign $J^{PC}$ quantum numbers.
Due to these issues in lattice QCD, it is not a problem that our calculations does not match theirs.

A recent study~\cite{Huang:2016rro} within  QCD SR for the exotic $0^{--}$ tetraquark with light quark content predicted a small mass,
$M_{tetra}=1.66\pm 0.14$ GeV. 
Therefore, the large mass difference should lead to a very small mixing between this light tetraquark state and the heavy exotic $0^{--}$ glueball.
However, one cannot avoid the discussion about possible mixing between the
exotic glueball states and the heavier tetraquarks
of the same exotic quantum number, if such heavy
tetraquarks exist. 
But we would like to point out that
 to our knowledge, all estimations within various models give
the value of the mass for the hidden-charm  tetraquarks to be around 4~GeV (see review~\cite{Chen:2016qju}), which is rather small in comparison to our glueball masses.
In principle, the exotic glueballs can also mix  with the hidden-charm hybrid 
which has the same quantum numbers.
The recent lattice calculation for $0^{+-}$ hybrid predicts the mass to be around 4.4~GeV~\cite{Liu:2012ze}.
Since there a large mass gap between the $0^{\pm-}$ glueballs and
the   exotic hadrons with hidden-charm, the mixing of the exotic glueballs with hidden-charm states is expected to be small.
In any case, the calculation of the mixing between different exotic states is very complicated due to contributions coming from
both the perturbative and nonperturbative sectors of QCD and
such kinds of studies are out of the scope of the present work.

The decay of the three-gluon state to the hadrons
is suppressed by the large power of the strong coupling  at
the virtuality of the glueball's gluons $Q^2\sim 4$~GeV$^2$,
where we assume that the gluons carry equal momenta.
One of the allowed channels includes charmonium in the final state.
In particular, 
we consider the S-wave decay of glueball $G(0^{--})\to f_1(1285)+J/\Psi$  to be the most preferable
 due to the large glueball mass and the small widths of the final particles.
Additionally, this channel could be enhanced by the decay of the hidden charm tetraquark.
Therefore, charmonium data could be a good place to search for experimental evidence of exotic glueballs.

\section{Acknowledgment}

We would like to thank M. Elbistan, S. Mikhailov, P. Gandini
and C. Halcrow
for stimulating discussions and useful remarks.
This work has been supported by the National Natural Science Foundation of China (Grants No. 11575254 and 11650110431),
Chinese Academy of Sciences President's International Fellowship Initiative
(Grant No. 2013T2J0011 and 2016PM053), the Japan Society for the Promotion
of Science (Grant No.S16019).
The work by H.J.L. was supported by the Basic Science Research Program
through the National Research Foundation of Korea (NRF) funded by Ministry
of Education under Grants No. 2016R1D1A1A09920078.
The work of  A.~P. and V.~K. has been also supported  by
the Russian Foundation for Basic Research under Grants No.\ 15-52-04023
and by the Belarusian Republican Foundation for Basic Research under Grants No.\ F15RM-072
respectively.


\end{document}